\documentclass[10pt,letterpaper,onecolumn]{article}
\usepackage{amsmath,amssymb,graphicx,cite,url,float,dsfont}
\usepackage[top=0.85in,bottom=0.85in,left=1.4in,right=1.4in,footskip=0.45in]{geometry}
\begin{document}
\begin{flushleft}
{\Large\textbf\newline{Secondary complementary balancing compressive imaging with a free-space balanced amplified photodetector}}

\bigskip
Wen-Kai Yu\textsuperscript{*},
Ying Yang,
Jin-Rui Liu,
Ning Wei,
Shuo-Fei Wang
\\
\bigskip
Center for Quantum Technology Research and Key Laboratory of Advanced Optoelectronic Quantum Architecture and Measurement of Ministry of Education, School of Physics, Beijing Institute of Technology, Beijing 100081, China\\
* yuwenkai@bit.edu.cn

\end{flushleft}

\date{}


\noindent\textbf{Abstract:} Single-pixel imaging (SPI) has attracted widespread attention because it generally uses a non-pixelated photodetector and a digital micromirror device (DMD) to acquire the object image. Since the modulated patterns seen from two reflection directions of the DMD are naturally complementary, one can apply complementary balanced measurements to greatly improve the measurement signal-to-noise ratio and reconstruction quality. However, the balance between two reflection arms significantly determines the quality of differential measurements. In this work, we propose and demonstrate a simple secondary complementary balancing mechanism to minimize the impact of the imbalance on the imaging system. In our SPI setup, we used a silicon free-space balanced amplified photodetector with 5~mm active diameter which could directly output the difference between two optical input signals in two reflection arms. Both simulation and experimental results have demonstrated that the use of secondary complementary balancing can result in a better cancellation of direct current components of measurements and a better image restoration quality.

\section{Introduction}
Single-pixel imaging (SPI) can acquire the object information from sequential single-pixel measurements of the superposition of the modulated patterns and the object. The single-pixel detection modality offers more possibilities for the wavelengths where pixelated detectors are technically unavailable or too expensive, such as x-ray, infrared, and terahertz wavelengths. This promising indirect imaging technique \cite{Duarte2008,Edgar2019,Gibson2020}, ever since it was proposed, has been intensively studied \cite{Shapiro2008,Katz2009} and provided countless novel image measurement ideas in science and engineering fields, including quantum entanglement \cite{Yang2020}, polarimetric imaging \cite{Duran2012}, three-dimensional tracking \cite{ZBZhang2020}, hyperspectral imaging \cite{Intes2017}, fluorescence microscopy \cite{Studer2012}, medical imaging \cite{Graff2015}, compressive holography \cite{Wu2021}, imaging through scattering media \cite{Duran2015}, optical encryption \cite{YuAO2019}, etc. Most of them were motivated by compressed sensing (CS) \cite{Donoho1992,Donoho2006,CBLi2010} and ghost imaging \cite{Gatti2004,Ferri2010} algorithms.

In SPI setups, the spatial modulation or structured illumination can be performed by using fast spatial light modulators (SLMs) such as modern digital micromirror device (DMD) with a fast switching frequency up to 22~kHz. To our knowledge, the modulation speed can be further increased by applying structured illumination with LED arrays \cite{Xu2018}, but at low spatial resolution. According to the nature of SPI, a lot of patterns are needed for per single-frame image acquisition. Given that the response frequency of the single-pixel detector is much faster than the modulation speed of the used SLM, the latter and the number of measurements place an upper bound to the acquisition speed of the system. The modulation speed is limited by devices and technologies, and is too hard to be increased. Any little increase in modulation speed comes at a huge cost. Therefore, to achieve imaging speed beyond the regular limit, many efforts are focused on improving the sampling efficiency by optimizing the ordering of the Hadamard basis patterns \cite{Sun2017,WKYu2019,WKYuorigami2019,Vaz2020} or using Fourier basis patterns \cite{Zhang2015}. Apart from these deterministic patterns, random patterns are also very popular in conventional SPI due to their simplicity and easy implementation. However, no matter which kind of the above methods is adopted, the measurements are susceptible to unfavorable ambient light, and the non-negativity of the patterns will produce a direct current (DC) component in the measured signal that cannot be ignored, leading to the degradation of reconstruction quality. Furthermore, considering the overlap between each pattern and the object part is basically random, the system will lose half flux of photons on average.

As we all know, each micromirror on the widely used DMD can rotate about a hinge and orientates $\pm12^\circ$ with respect to its normal direction, corresponding to ``on'' (1) and ``off'' (0) binary states. Thus, the light will be reflected in two directions depending on the modulation matrices loaded onto the DMD. When we look at the DMD along the opposite of these two reflection directions, only the mirrors in the ``on'' (``off'') state will reflect the light while the ones in the ``off'' (``on'') state will be dark, thus in two reflection arms during each modulation, we will see two patterns that are exactly complementary. Given this, double-arm dual-pixel complementary differential (or positive-negative) measurement was proposed and proven to improve the measurement signal-to-noise ratio (SNR) and to make full use of all photon flux by setting two single-pixel detectors in both reflection directions \cite{WKYu2014,Radwell2014}. One could also make the DMD modulate one pattern immediately followed by its inverse (complementary) pattern and take the difference between adjacent single-pixel measurements to achieve a similar effect \cite{WKYu3D2015,YuTracking2015,Yu2016}. We call this method single-arm single-pixel complementary differential measurement. By using the above two methods, the DC term of the measured values can be eliminated, while the alternating current (AC) term that responds to the fluctuations of the measurements can be retained to recover the object image of interest. Moreover, the sampling ratio can be shortened to a much lower level than the traditional SPI when getting the same image quality, and the immunity to environmental light and temperature drift of light source can also be enhanced. Later, the complementary differential measurement was performed by directly using a single free-space balanced detector \cite{Soldevila2016,Denk2019}, which is integrated with two single-pixel detectors and can improve the signal digitization process. By subtracting the two optical input signals from each other, the cancellation of common mode noise can be realized. Actually, the use of balanced measurements in optical experiments can be traced back to the late 1960s \cite{Carleton1968}. In above SPI schemes with balanced detection, the light in both reflection directions of the DMD was coupled to two optical fibers by using two identical fiber collimating elements \cite{Soldevila2016} or refocused onto two well-matched photodiodes of the balanced detector by attaching a pair of lenses onto these two photodiodes \cite{Denk2019}. The use of two optical fibers makes the setup flexible, but at the expense of optical coupling (collection) efficiency. Both dual-pixel detection and balanced detection belong to double-arm complementary measurement modality. However, it is inevitable that there will be differences in optical collection between two arms, even with the use of industrialized opto-mechanical system design. The balance between two reflection arms becomes very crucial and directly determines the suppression level of the DC component and the image reconstruction quality.

In this work, we propose a simple secondary complementary balancing mechanism on the basis of a complementary SPI scheme with balanced detection. The impact of the optical imbalance in double reflection arms of the DMD will be minimized. The DC term of the measured values can be eliminated to the maximum extent, and the fluctuations of the measurements that are really useful for image reconstruction will be amplified, thus leading to the increase of the measurement SNR and accuracy. Besides, the full dynamic range of the detector can be used to record the positive-negative fluctuations. The feasibility and superiority of this method will be demonstrated through both numerical simulations and optical experiments.

\section{Methods}

In SPI, the single-pixel detector records the inner product (superposition) of the object image $O(c,d)$ and each modulated pattern $P_i(c,d)$, denoted as $Y_i=\sum_{c=1}^p\sum_{d=1}^qP_i(c,d)O(c,d)$. The modulated patterns are of the same pixel-size $p\times q=N$ with the object image. As specified by the Nyquist-Shannon criterion, we need to perform full sampling, i.e., we should make the number of measurements equal to $N$. For instance, the Hadamard and Fourier patterns are two classic deterministic orthogonal bases which are widely used for full sampling. To reduce the sampling ratio, one can use CS theory by exploiting the sparsity of the object image. The modulated patterns used in compressive sampling scheme are generally 0-1 random, so the overlap between each pattern and the object part will also be random, causing a loss of half flux of photons on average. Besides, the non-negativity of each pattern will generate a DC term in each measured value. Note that one pattern will present as a complementary pair in two reflection directions of the DMD, we proposed a complementary differential measurement technology in our previous work \cite{WKYu2014}, where two reflection arms are sampled simultaneously by two single-pixel detectors. By this way, the common mode noise can be directly eliminated, the measurement SNR is increased, and the entire flux of photons can be fully utilized. The two single-pixel detectors can also be replaced by a free-space large-area balanced amplified photodetector (BAP) \cite{Soldevila2016,Denk2019}, which can directly output the voltage that is proportional to the difference between the photocurrents in two arms, but without additional synchronization burden. However, we need to set multiple lenses even with multiple plane mirrors in both reflection arms for the sake of symmetrical optical path design. Even with industrialized opto-mechanical system design, we cannot strictly guarantee that these two light paths are completely the same and the collected spots as well as the collection efficiencies on two photodetectors are the same. Given this, it is hard to ensure that the common mode noise can be eliminated to almost zero. To address this issue, we design a simple secondary complementary balancing method here.

Assume $Y_i^+$ and $Y_i^-$ are two single-pixel signals recorded in two reflection directions of the DMD (corresponding to $+12^\circ$ and $-12^\circ$ orientations of micromirrors, respectively), and can be expressed as

\begin{equation}
Y_i^+=\sum_{c}\sum_{d}P_i^+(c,d)O(c,d),
\end{equation}
\begin{equation}
Y_i^-=\sum_{c}\sum_{d}P_i^-(c,d)O(c,d),
\end{equation}

\noindent where the $i$th complementary random pattern pair ($P_i^+$ and $P_i^-$) satisfies $P_i^++P_i^-=\mathds{1}$. Here $\mathds{1}$ denotes a matrix of the same size of $P_i^+$ and $P_i^-$ consisting of all ones.

If the object image $O$ is flattened into a column $x$ of $N\times1$, then each modulated pattern can be reshaped into a row vector of $1\times N$, and $M$ such patterns will form a measurement matrix $A$ of $M\times N$. The single-pixel values will constitute a sequence of $M\times1$ which can be written as $y=Ax+e$, where $e$ of the same size $M\times1$ stands for the stochastic noise. Generally, we can find the sparse representations of natural object images in some invertible (e.g., orthogonal or approximately orthogonal) or redundant bases $\Psi$, i.e., $x=\Psi x'$. Then, we will have $y=A\Psi x'+e$. In the complementary measurement scheme \cite{WKYu2014}, the single-pixel signals measured by two photodetectors can be written as

\begin{equation}
y^+=A^+\Psi x'+e^+,
\end{equation}
\begin{equation}
y^-=A^-\Psi x'+e^-,
\end{equation}

\noindent where $A^+$ and $A^-$ are complementary, all consisting of 0 and 1, $e^+$ and $e^-$ denote the noise in two reflection arms. Their differential signal can be written as

\begin{equation}
y^+-y^-=\hat{A}\Psi x'+e^+-e^-,
\end{equation}

\noindent where $\hat{A}=A^+-A^-$. In the ideal model, we expect the two-arm optical paths to be completely symmetrical, with no luminous flux loss, no difference in propagation distance and collection efficiency, and the environmental noise in two reflection arms is of the same magnitude (i.e., independent and identically distributed). But in actual measurement, it is difficult for us to acquire the above expected conditions.

To solve this problem, we assume that the single-pixel values measured in the two arms have a linear transformation with the expected values, then we will have

\begin{equation}
y'^+=k^+(A^+\Psi x'+e^+)+b^+,
\label{eqy1positive}
\end{equation}
\begin{equation}
y'^-=k^-(A^-\Psi x'+e^-)+b^-,
\label{eqy1negative}
\end{equation}

\noindent where $k^+$ and $k^-$ stand for the proportionality coefficients, $b^+$ and $b^-$ are constant bias. If we sequentially encode a complementary pattern onto the DMD, two signals received by two photodiodes can be written as

\begin{equation}
y''^+=k^+(A^-\Psi x'+e^+)+b^+,
\label{eqy2positive}
\end{equation}
\begin{equation}
y''^-=k^-(A^+\Psi x'+e^-)+b^-,
\label{eqy2negative}
\end{equation}

\noindent where the proportionality coefficients and constant bias keep unchanged, but the modulated patterns seen in two detection directions are changed to their inverse ones.

Since the BAP will automatically output the differential signal of the two arms, the full dynamic range of the BAP can be used to record the positive-negative fluctuations. Then, we will get two differential measurements for a pair of complementary modulation patterns:

\begin{equation}
B_1=y'^+-y'^-=(k^+A^+-k^-A^-)\Psi x'+k^+e^+-k^-e^-+b^+-b^-,
\end{equation}
\begin{equation}
B_2=y''^+-y''^-=(k^+A^--k^-A^+)\Psi x'+k^+e^+-k^-e^-+b^+-b^-.
\end{equation}

It is not difficult to find that there are many mutually canceling and merging terms in the mathematical expressions of these two differential signals. Thus, we make the difference between these two signals and get the following equation

\begin{equation}
B_1-B_2=[(k^+(A^+-A^-)-k^-(A^--A^+)]\Psi x'=\hat{A}[(k^++k^-)\Psi x']=k'\hat{A}\Psi x',
\end{equation}

\noindent where $k'=k^++k^-$. It can be seen from this equation that the environmental noise is eliminated to the greatest extent, causing to the increase of the measurement SNR and accuracy, and the reconstructed image will be $k'$ times of the original one, leading to an increase in image contrast but not affecting the actual reconstructed image content. More importantly, by this means, the impact of the optical imbalance in double detection arms is minimized. In addition, it is worth noting that the differential operations of the first layer are achieved on the BAP device, while the differential operation of the second layer is completed by data post-processing. Given this, we name this method as secondary complementary balancing strategy.

\section{Simulation}

\begin{figure}[htbp]
\centering\includegraphics[width=10cm]{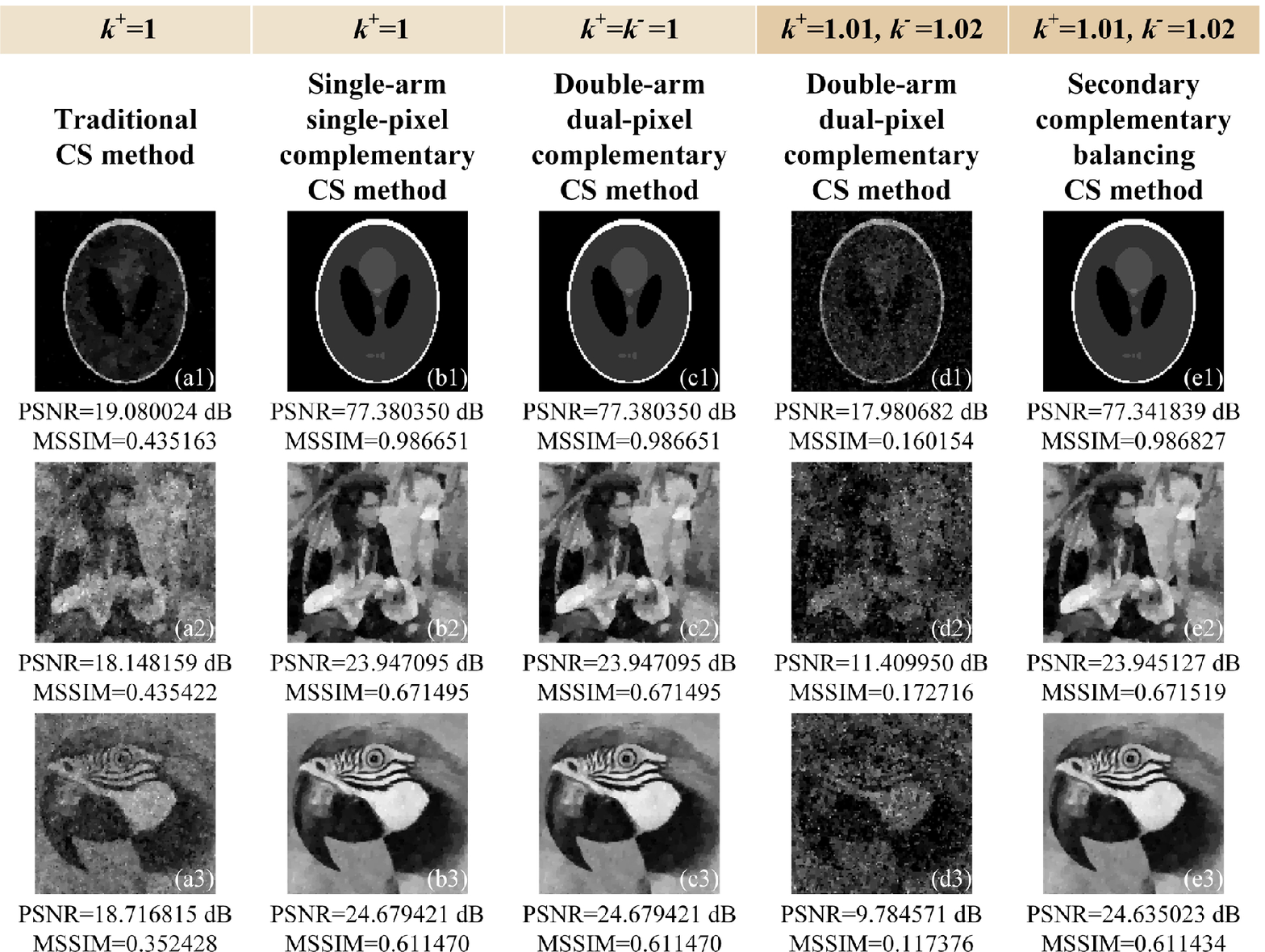}
\caption{Simulation results of traditional CS method without complementary measurements (\textbf{a1--a3}), single-arm single-pixel complementary CS method (\textbf{b1--b3}), double-arm dual-pixel complementary CS method in both ideal situation (\textbf{c1--c3}) and imbalanced situation (\textbf{d1--d3}), and secondary complementary balancing CS method in imbalanced case (\textbf{e1--e3}), respectively. The pixel-sizes of reconstructed images are $128\times128$ and the total sampling ratios are all 25\%.}
\label{fig1}
\end{figure}

We first made a comparison between traditional CS method without complementary measurements, single-arm single-pixel complementary CS method, double-arm dual-pixel complementary CS method (in both ideal and imbalanced situations), secondary complementary balancing CS method (only in imbalanced situation), and presented their simulation results in Figure~\ref{fig1}. All these four methods applied a total variation minimization (TVAL3) solver \cite{CBLi2010} to solve the ill-posed underdetermined linear problem (where the number of equations or measurements is less than the number of unknowns). We chose three grayscale standard test images (i.e., head phantom, man and parrot) of $128\times128$ pixels as our original images. To make a fair comparison, the total sampling ratios applied for these four methods are all fixed to 25\%. To acquire a quantitative measure of the reconstruction performance, the peak signal-to-noise ratio (PSNR) and mean structural similarity (MSSIM) \cite{Simoncelli2004} are used here as a figure of merit. The PSNR can evaluate the pixel error between an original image $U_o$ and a reconstructed image $\tilde U$, defined as $\textrm{PSNR}=10\log(255^2/\textrm{MSE})$, where $\textrm{MSE}=\frac{1}{pq}\sum\nolimits_{c,d=1}^{p,q}[U_o(c,d)-\tilde U(c,d)]^2$. But it does not consider the visual recognition perception characteristics of the human eye. Instead, the MSSIM is a full reference metric, which is based on the assumption that the human eye will extract the structural information when viewing the image. Naturally, the larger are the PSNR and MSSIM values, the better is the image quality.

\begin{figure}[htbp]
\centering\includegraphics[width=\textwidth]{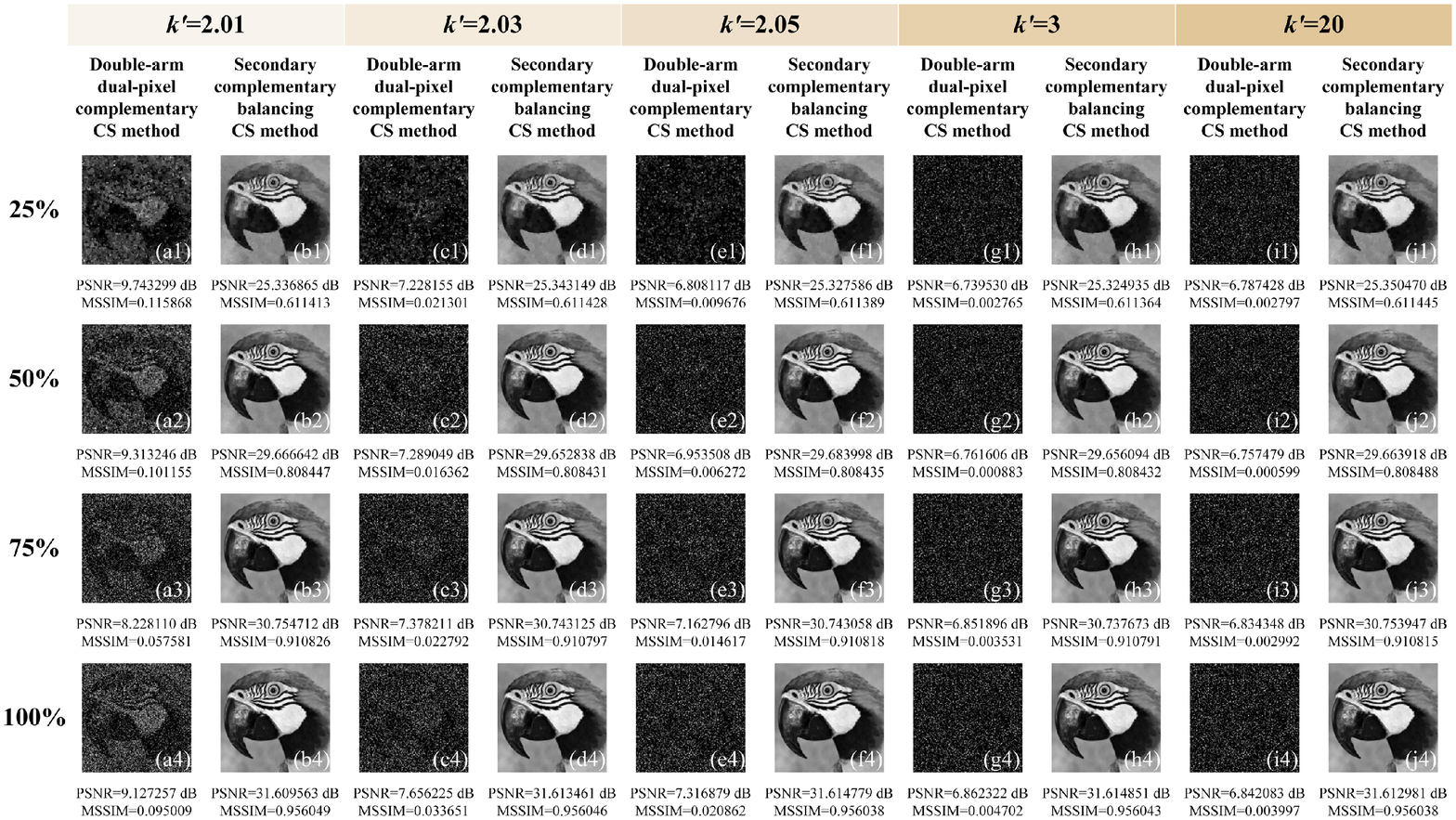}
\caption{Simulation results of the image parrot by using traditional double-arm dual-pixel complementary CS and secondary complementary balancing CS method, with different sampling rates 25\%, 50\%, 75\% and 100\% (from top to bottom) and different imbalance coefficients $k'=2.01$, 2.03, 2.05, 3 and 20 (from left to right).}
\label{fig2}
\end{figure}

As we can see from Figure~\ref{fig1}a1--b3 that single-arm single-pixel complementary CS method can definitely obtain a better performance than traditional CS approach when using the same total sampling ratio, and both only use a single-arm light path for detection, thus there is no optical imbalance problem. Utilizing the natural complementary property of the patterns seen in two reflection directions of the DMD, we can convert two successive complementary measurements (in single-arm) into two simultaneous measurements (in double-arm) and acquire the equivalent imaging results under ideal conditions (see Figure~\ref{fig1}c1--c3). Thus, both single-arm single-pixel complementary CS method and double-arm dual-pixel complementary CS method have the same total number of measurements. But in actual measurement, it is difficult for us to ensure that the optical paths of the two arms are absolutely symmetrical with balanced light intensities. Here we give an imbalanced case of $k^+=1.01$, $k^-=1.02$, $b^+=1$, $b^-=5$ as an example, the double-arm dual-pixel complementary CS method will present a sharp decline in its reconstruction performance (see Figure~\ref{fig1}d1--d3). As can be seen from this example, the image degradation problem caused by the asymmetry in two arms cannot be ignored. It is worth mentioning that $b^+$ and $b^-$ are constants which will not have much influence on image reconstruction according to the theory of CS, and can be eliminated by our secondary complementary balancing. From Figure~\ref{fig1}e1--e3, we can clearly see that in imbalanced situation our secondary complementary balancing CS method can acquire similar imaging performance with both single-arm single-pixel complementary CS method and double-arm dual-pixel complementary CS method (in ideal situation), and much better than that of double-arm dual-pixel complementary CS method in imbalanced situation (see Figure~\ref{fig1}d1--d3). Therefore, our secondary complementary balancing CS method can effectively eliminate the influence of optical imbalance. Besides, since the differential measurement function of the BAP is used, a set of complementary differential measurements is combined into one balanced measurement. Thus, the complementary modulation will not increase its total number of measurements compared with conventional complementary CS method.

\begin{figure}[htbp]
\centering\includegraphics[width=12cm]{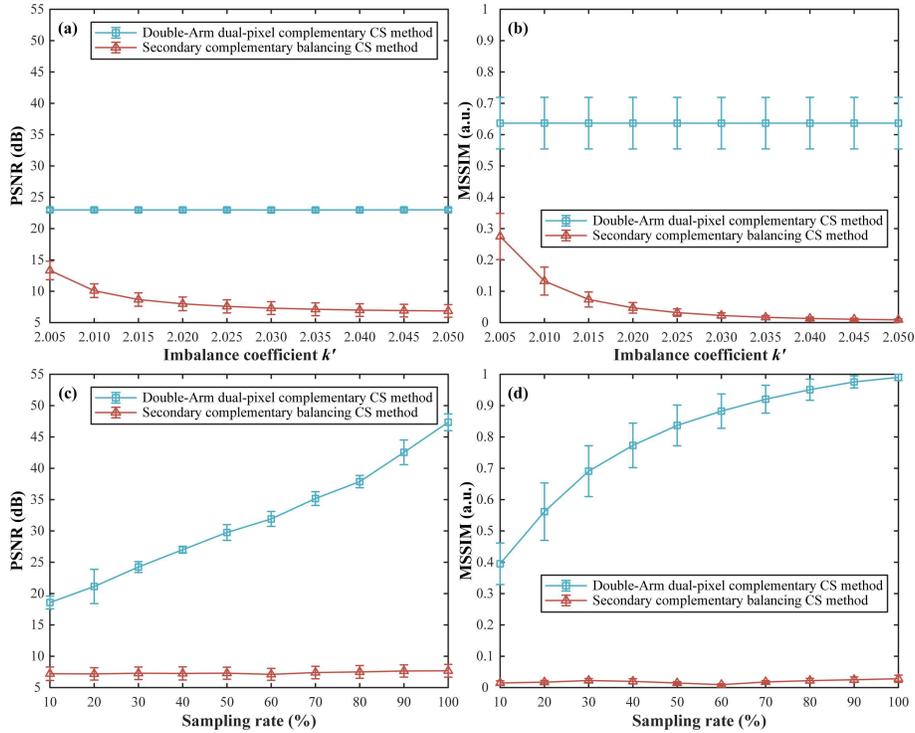}
\caption{Comparisons of reconstructed results of several general scenes by using double-arm dual-pixel complementary CS method and secondary complementary balancing CS method. (\textbf{a}) and (\textbf{b}) are the PSNR and MSSIM curves as a function of the imbalance coefficient $k'$, respectively. (\textbf{c}) and (\textbf{d}) are the PSNR and MSSIM curves as a function of the sampling rate, respectively. In these curves, each point is acquired by averaging PSNR or MSSIM results of five different complex grayscale object scenes, including Barbara, goldhill, sailboat, cablecar and pens, all of $128\times128$ pixels. The half height of each error bar indicates the standard deviation of each point.}
\label{fig3}
\end{figure}

Then, we changed the total sampling ratio and imbalance coefficient $k'$ to test the imaging performance. Without loss of generality, we directly set $k^+=1$ and assumed that the two reflection arms of the DMD have a difference of 1\%, 3\%, 5\%, 200\% and 1900\% in recorded light intensities, respectively, i.e., $k'=2.01$, 2.03, 2.05, 3 and 20. As the asymmetry increases, more and more cluttered points appear in the image reconstructed by traditional double-arm dual-pixel complementary CS method, as shown in Figure~\ref{fig2}a1--a4, c1--c4, e1--e4, g1--g4 and i1--i4. As we can see from Figure~\ref{fig2}b1--b4, d1--d4, f1--f4, h1--h4 and j1--j4 that our secondary complementary balancing CS method can well suppress the effects caused by the imbalance of the optical paths and acquire a better image quality even at low sampling ratios. For a more intuitive comparison, we further drew the variation curves of the PSNRs and MSSIMs of these two methods with the increase of the imbalance coefficients as well as the increase of the sampling rates, as shown in Figure~\ref{fig3}. These curves also demonstrate that our secondary complementary balancing CS method performs better than conventional double-arm dual-pixel complementary CS method.

\begin{figure}[htbp]
\centering\includegraphics[width=\textwidth]{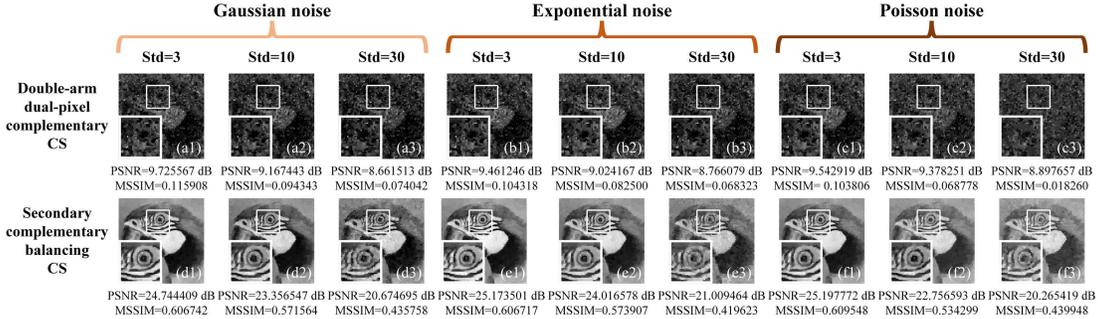}
\caption{Reconstructed results of $128\times128$ pixels with 25\% sampling rate under additive noise of three kinds of distributions. (\textbf{a1--a3}) and (\textbf{d1--d3}), (\textbf{b1--b3}) and (\textbf{e1--e3}), (\textbf{c1--c3}) and (\textbf{f1--f3}) are the results of double-arm dual-pixel complementary CS method and secondary complementary balancing CS method under the Gaussian, exponentially and Poisson distributed noise, respectively. Here $\textrm{Std}$ denotes the standard deviation of the noise distribution.}
\label{fig4}
\end{figure}

Next, we further investigated the effect of three kinds of additive noise on image quality. Here we chose the Gaussian, exponentially and Poisson distributed noise, and directly added them onto the measured values. The results were given in Figure~\ref{fig4}, where $\textrm{Std}$ stands for the standard deviation of the distribution which the additive noise follows. All images were recovered with 25\% sampling rate under the situation of $k^+=1$ and $k^-=1.01$. From Figure~\ref{fig4}a1--a3, b1--b3 and c1--c3, we could see that the optical imbalance has a significant degrading effect on the reconstructed images when using double-arm dual-pixel complementary CS method. While in Figure~\ref{fig4}d1--d3, e1--e3 and f1--f3, it could be seen that the secondary complementary balancing CS method has a good robustness against measurement noise, with the image details (see the enlarged images marked in the white square frames) being well preserved.

\section{Experiments}

\begin{figure}[htbp]
\centering\includegraphics[width=10cm]{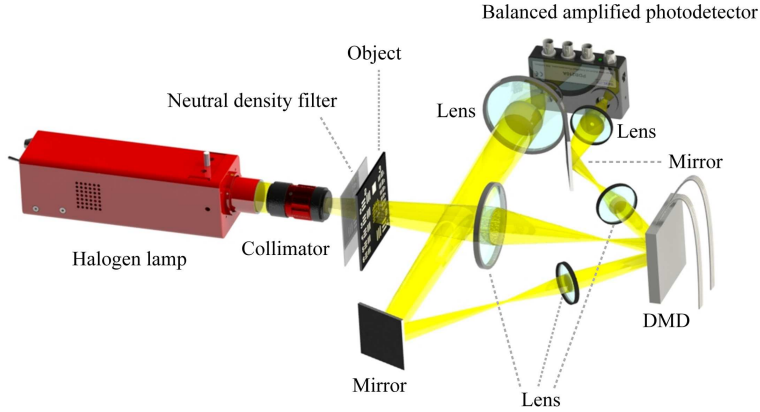}
\caption{Experimental setup of secondary complementary balancing compressive imaging. DMD: digital micromirror device.}
\label{fig5}
\end{figure}

Our experimental setup is given in Figure~\ref{fig5}: the thermal light (ranging from 360~nm to 2600~nm) emitted from a halogen lamp is collimated by a collimator and attenuated by some 2~inch $\times$ 2~inch neutral density filters to the ultra-weak light level, and then illuminates an object (a negative 1951 USAF resolution test chart of 3~inch $\times$ 3~inch). The transmitted light is projected vertically via an imaging lens onto a 0.7~inch DMD consisting of $768\times1024$ micromirrors (i.e., pixels), which is encoded with pre-prepared complementary patterns. The reflected light of the DMD is emitted at angles of $\pm24^\circ$ with respect to the normal of the DMD's work plane. The light from two reflection arms separately passes through two sets of lenses and a mirror, and is collected by a BAP (Thorlabs PDB210A/M), which will directly output the differential signal of two photodiodes' photocurrents. The output voltage signal will be fully recorded by a mixed signal oscilloscope (Tektronix MSO64 6-BW-4000).

\begin{figure}[htbp]
\centering\includegraphics[width=\textwidth]{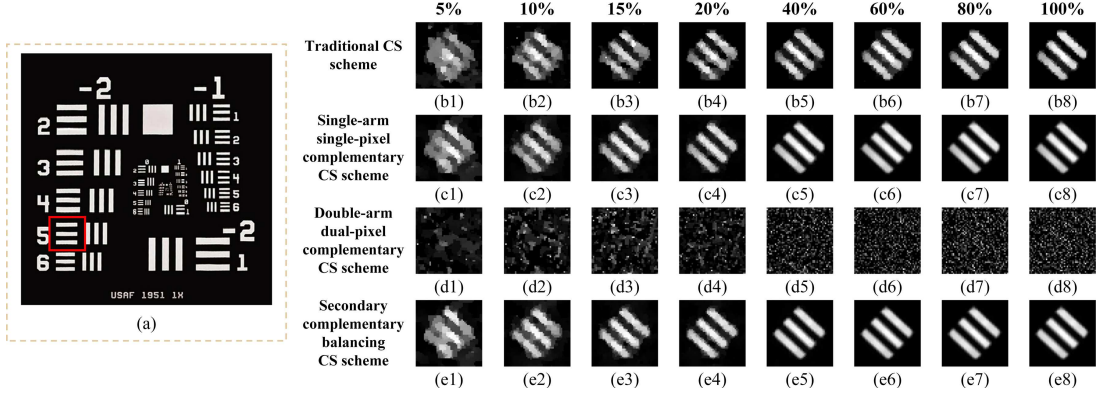}
\caption{Experimental results of three CS schemes using different sampling rates (5\%, 10\%, 15\%, 20\%, 40\%, 60\%, 80\% and 100\% from left to right) in the imbalanced case. (\textbf{a}) The stripe pattern marked in the red square of the negative 1951 USAF resolution test chart is treated as the original object. (\textbf{b1--b8}), (\textbf{c1--c8}), (\textbf{d1--d8}) and (\textbf{e1--e8}) are the results of traditional CS scheme without complementary measurements, single-arm single-pixel complementary measurement scheme, double-arm dual-pixel complementary measurement scheme, and secondary complementary balancing CS scheme, respectively.}
\label{fig6}
\end{figure}

In experiment, we also tested the practical performance of traditional CS scheme without complementary measurements, single-arm single-pixel complementary CS scheme, double-arm dual-pixel complementary CS scheme, and secondary complementary balancing CS scheme. Here, the complementary patterns were generated from the random patterns used in the first method. Since the BAP has two additional interfaces, which can independently output the photocurrents of two photodetectors, traditional CS scheme without complementary measurements can be performed by using the light intensities (in one reflection direction) collected by one photodetector of the BAP. Since the patterns seen in two reflection directions of the DMD were exactly complementary, we made a difference between the photocurrents detected by two photodetectors. By this means, traditional double-arm dual-pixel complementary measurement scheme \cite{WKYu2014,Radwell2014} was realized. Then, we made the DMD modulate one random pattern immediately followed by its inverse/complementary one, and also used the photocurrents that were recorded in one reflection arm to realize single-arm single-pixel complementary measurement \cite{WKYu3D2015,YuTracking2015,Yu2016}. While in our secondary complementary balancing CS scheme, the voltage that was proportional to the difference between the photocurrents in two arms was used directly for image reconstruction, and the complementary modulation was also applied. The patterns encoded onto the DMD were of $48\times48$ pixel-units, and each of which would occupy $12\times12=144$ micromirrors of the DMD, i.e., a total of $576\times576$ micromirrors were actually involved in the optical modulation. The experimental results of these three schemes under different sampling ratios (ranging from 5\% to 100\%) were presented in Figure~\ref{fig6}. As shown in Figure~\ref{fig6}a, the red square marked in the negative 1951 USAF resolution test chart was chosen here as the original image, whose stripes are separated by 1.26~mm on the resolution test chart. It can be clearly seen that the imaging quality of traditional CS scheme, single-arm single-pixel complementary measurement scheme and secondary complementary balancing CS scheme improves with the increase of the sampling ratio. Due to the optical asymmetry, the double-arm dual-pixel complementary measurement scheme could not acquire the image quality improvement as the sampling ratio increases. The reconstruction qualities of our secondary complementary balancing CS scheme (see Figure~\ref{fig6}e1--e8) are much better than those of traditional CS scheme (see Figure~\ref{fig6}b1--b8), traditional double-arm dual-pixel complementary measurement scheme (see Figure~\ref{fig6}d1--d8), and similar to those of single-arm single-pixel complementary measurement scheme (see Figure~\ref{fig6}c1--c8), at any sampling ratios. In addition, when the sampling rate reached 40\%, the image quality of this proposed method was almost the same with that of traditional CS scheme without complementary measurements under full sampling condition. These results are consistent with the simulation results. It is worth mentioning that there is in fact no optical imbalance in the single-arm single-pixel complementary measurement scheme, and its experimental results were listed here only for reference because it also uses complementary modulation. Therefore, our secondary complementary balancing CS scheme can effectively eliminate the image degradation problem caused by optical imbalance during dual-arm measurements.

\begin{figure}[htbp]
\centering\includegraphics[width=9cm]{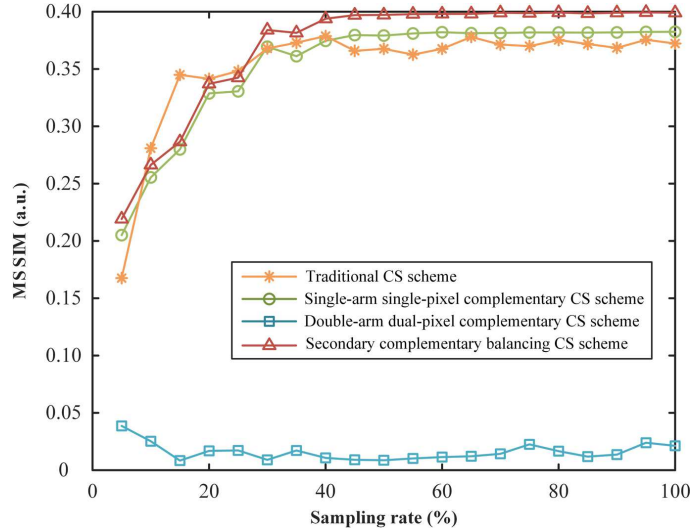}
\caption{Curves of MSSIM values of traditional CS scheme, single-arm single-pixel complementary CS scheme, double-arm dual-pixel complementary CS scheme and secondary complementary balancing CS scheme as a function of the sampling rate.}
\label{fig7}
\end{figure}

Next, we further analyzed the performance of the above four kinds of schemes, by plotting the MSSIM curves as a function of the sampling ratio as shown in Figure~\ref{fig7}. From the curves we could see that when the sampling rate is higher than 40\%, our method outperforms the others. Although there is only one collecting light path (without optical imbalance problem) in both traditional CS scheme and single-arm single-pixel complementary CS scheme, their reconstruction qualities are still slightly lower than those of our secondary complementary balancing CS scheme, especially at high sampling rates. This is because our secondary complementary balancing strategy can not only eliminate optical imbalance, but also suppress measurement noise by noise balancing. It should be noted here that in the curves shown in Figure~\ref{fig7}, the MSSIM values of the traditional CS scheme present some fluctuation at low sampling rates, which is actually an estimation deviation caused by the fact that the MSSIM as a full-parameter image quality metric relies too much on pixel values for image evaluation. From Figure~\ref{fig6}, we could clearly observe that the visibility of the reconstructed images via traditional CS scheme is always lower than those of the other three schemes at any sampling rate.

\begin{figure}[htbp]
\centering\includegraphics[width=10cm]{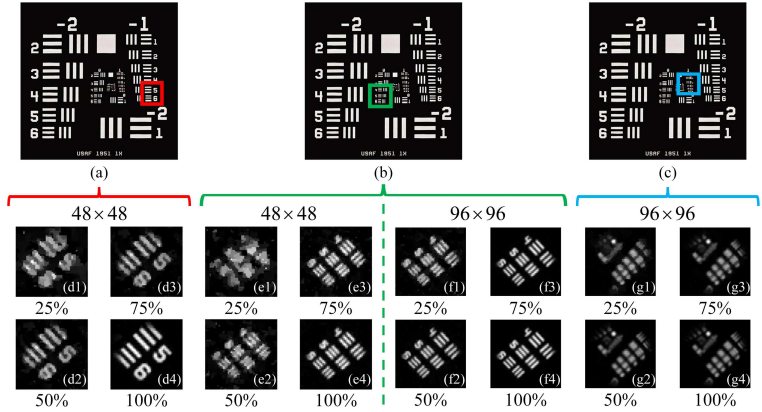}
\caption{Reconstructed images of secondary complementary balancing CS scheme for different imaging regions of the resolution test chart with different sampling rates. (\textbf{a--c}) give three imaging regions in the resolution test chart. (\textbf{d1--d4}) are the recovered images of the object region (\textbf{a}) by using modulated patterns of $48\times48$ pixel-units; (\textbf{e1--e4}) and (\textbf{f1--f4}) are the results of the object region (\textbf{b}) by using patterns of $48\times48$ and $96\times96$ pixel-units, respectively; and (\textbf{g1--g4}) are the reconstructed images of the object region (\textbf{c}) by applying patterns of $96\times96$ pixel-units.}
\label{fig8}
\end{figure}

Then, we further analyzed the imaging performance of our secondary complementary balancing CS scheme in different spatial resolutions and different imaging regions of the resolution test chart, as shown in Figure~\ref{fig8}. Figure~\ref{fig8}d1--d4 are the $48\times48$-pixel-unit reconstructed images of the original object marked in the red square of Figure~\ref{fig8}a by using different sampling ratios. Note that the red square locates in Group -1, the parallel lines for Elements 5 and 6 are 3149.80~$\mu$m and 2806.16~$\mu$m long, and 629.96~$\mu$m and 561.23~$\mu$m wide with space 629.96~$\mu$m and 561.23~$\mu$m wide between the parallel lines, respectively. Then, we changed the imaging region to the green square (see Figure~\ref{fig8}b) and used patterns of $48\times48$ and $96\times96$ pixel-units. The corresponding results are provided in Figure~\ref{fig8}e1--e4 and f1--f4. Since the green square is in Group 0, the parallel lines for Elements 4, 5 and 6 are 1767.77~$\mu$m, 1574.90~$\mu$m and 1403.08~$\mu$m long, and 353.55~$\mu$m, 314.98~$\mu$m and 280.62~$\mu$m wide, separated by equal spaces of 353.55~$\mu$m, 314.98~$\mu$m and 280.62~$\mu$m wide, respectively. For the pixel-size of $48\times48$, when the sampling ratio is higher than 75\%, the image quality improvement tends to saturate, while for $96\times96$ pixel-units, when the sampling ratio is higher than 50\%, the image quality increase tends to saturate. It can be seen that as the spatial resolution of patterns increases, the reconstructed images will be clearer. Therefore, the pixel resolution of the patterns is an important parameter affecting the imaging quality, but the increase of the pixel resolution will also bring in an increase in computing time and memory consumption (so a trade-off needs to be made in sampling process). The stripe pattern (involving Groups 1--3) chosen in Figure~\ref{fig8}c are more refined, and the parallel lines in the reconstructed images (see Figure~\ref{fig8}g1--g4) of $96\times96$ pixel-units are difficult to distinguish.

It is worth mentioning that the random patterns were mainly used in this paper for illustration, but the modulated patterns are not limited to this. Since the modulated patterns in balanced detection scheme need to be differentiated, the measurement matrix is positive-negative distributed, Thus, the Hadamard patterns consisting of $\pm1$ will be another option. For example, the popular optimization sorting (including Russian dolls sorting \cite{Sun2017}, cake-cutting sorting \cite{WKYu2019}, origami sorting \cite{WKYuorigami2019}) of the Hadamard basis in recent years can be also used in this secondary complementary balancing scheme to further reduce the sampling ratios required for acquiring good image quality. Since this is not the focus of this paper, it will not be described in detail in this paper.

\section{Conclusion}

In this paper, a secondary complementary balancing mechanism is proposed and combined with CS to eliminate the effect of optical asymmetry (imbalance) on image reconstruction degradation that exists in double-arm balanced (differential) detection of complementary SPI scheme. Here, we use a free-space BAP to perform balanced measurements. By making the DMD modulate a pattern immediately followed by its complementary one, we can achieve a good cancellation of DC components of detected values as well as a high-quality image reconstruction. The use of secondary complementary balancing compressive measurement can improve the measurement SNR and accuracy by amplifying the fluctuating portions of the measured values and making full use of entire dynamic range of BAP to record these positive-negative fluctuations. Both numerical simulation and optical experimental results have demonstrated the feasibility and superiority of this technique. The influence of on image quality have been investigated. We believe that this mechanism will be very helpful in double-arm complementary SPI and may offer many benefits especially for balanced detection.

\section*{\textsf{Funding}}
Beijing Natural Science Foundation (4222016); Civil Space Project of China (D040301); Youth Talent Promotion Project of the Beijing Association for Science and Technology (none).

\section*{\textsf{Disclosures}}
The authors declare that there are no conflicts of interest related to this article.

\bigskip





\end{document}